\documentclass[conference]{IEEEtran}
\IEEEoverridecommandlockouts

\usepackage{cite}
\usepackage{amsmath,amssymb,amsfonts}
\usepackage{algorithm}
\usepackage{algpseudocode}
\usepackage{graphicx}
\usepackage{booktabs}
\usepackage{xcolor}
\usepackage{tikz}
\usepackage{pgfplots}
\pgfplotsset{compat=1.18}
\usetikzlibrary{arrows.meta,positioning,shapes.geometric,calc,fit}
\usepackage{listings}
\usepackage{url}
\usepackage[hidelinks]{hyperref}

\newcommand{\code}[1]{\texttt{#1}}

\begin{document}

\title{ResumeShield: Channel Separation and an Open\\
Benchmark for Indirect Prompt Injection\\ in AI Resume Screening}

\author{\IEEEauthorblockN{Jay Barach}
\IEEEauthorblockA{\textit{Independent Researcher}\\
\url{https://github.com/jbarach2012/ResumeShield}}}

\maketitle

\begin{abstract}
An AI resume screener reads a document supplied by the person it is evaluating,
which inverts the usual trust relationship between an assessor and the material it
assesses. Candidates exploit this by concealing instructions inside a resume using
white text, a zero font size, hidden elements, markup comments, document metadata,
or zero width characters. A human reviewer sees nothing, while a naive extraction
pipeline places the concealed text directly into the model prompt, where it is read
as an instruction and obeyed. This is indirect prompt injection, listed as
LLM01:2025 by OWASP, and recent measurement work reports it in roughly one percent
of resumes in a production screening corpus. We present ResumeShield, an
open-source defense and benchmark. The defense combines three filtering stages with
a fourth architectural stage that places candidate content in an explicitly fenced
data channel that the operator's trusted instructions declare to be inert. The
benchmark builds a seeded synthetic corpus spanning nine concealment techniques and
two payload families, one using documented phrasings and one modeling an adaptive
attacker who paraphrases around the filter, and it scores an attack as successful
only when the screening outcome actually changes. On a corpus of 104 documents the
naive pipeline is manipulated in every injected case while the defended pipeline is
never manipulated. Detection reaches a precision of 1.000 and a recall of 0.944 with
no false positives on clean resumes. An ablation isolates the contribution of each
stage and shows that channel separation alone removes all measured attack success,
whereas the complete filtering stack without separation still leaves 16.7 percent of
attacks effective. We also identify a concealment dilemma: every payload that
evaded detection was one the attacker left visible, which surrenders the
invisibility that motivates the attack. ResumeShield is released under the Apache
2.0 license with synthetic data only.
\end{abstract}


\section{Introduction}
Large language models increasingly perform the first pass of candidate screening,
reading a resume against a job description and returning a score, a ranking, or a
verdict. The arrangement has a structural weakness that distinguishes it from most
document processing tasks: the subject of the evaluation supplies the input to the
evaluator. If any portion of that input reaches the model as an instruction rather
than as material to be judged, the candidate can steer the outcome.

This is indirect prompt injection. Greshake and colleagues established the general
form of the attack, showing that applications which blend retrieved content with
operator instructions blur the boundary between data and instructions and thereby
expose a remote attack surface \cite{greshake}. The OWASP project for large language
model applications lists prompt injection first among its risks, as LLM01:2025
\cite{owaspllm}. The hiring setting gives the attack a specific and unusually clean
shape, because the attacker controls an entire document that a human may open and
inspect. The payload must therefore satisfy two constraints at once: it must survive
text extraction so the model receives it, and it must remain invisible to a reviewer
who looks at the rendered page. That pair of constraints produces the concealment
techniques observed in practice, including text colored to match the background, a
font size of zero, elements positioned off the canvas, instructions parked in markup
comments or document metadata, and zero width characters interleaved through a
sentence.

The problem is not hypothetical. A recent measurement study over approximately two
hundred thousand real resumes collected by a commercial screening vendor reports that
around one percent contain hidden prompt injections, that the prevalence rose over
the preceding one to two years, and that color based hiding, in which text is drawn
in the background color, is among the recurring techniques \cite{measuring}. Separate
controlled work finds that injected self promotional text reliably improves rankings
when candidate quality is similar and few applicants inject, and that it can allow a
weaker candidate to outrank a stronger one \cite{single_multi}. Related work on
ranking systems shows the same decision hijacking effect when documents compete
against one another \cite{ranking_blind}.

Defenses fall into two broad families. The first is detection and filtering, in
which the pipeline searches the candidate document for instruction like content and
removes or rejects it. The second is structural, in which the system prevents
untrusted content from occupying the instruction channel at all. The most convincing
structural results to date, including structured queries \cite{struq}, preference
optimized alignment \cite{secalign}, and instruction hierarchies \cite{hierarchy},
achieve their separation by training or fine tuning the model, which presumes access
that most application teams do not have. The practical question for an operator who
consumes a hosted model through an API is therefore what separation is achievable at
the application layer, and how much it is worth relative to filtering.

This paper answers that question for the resume screening setting. We present
ResumeShield, an open-source system that combines filtering with application layer
channel separation, together with a benchmark that measures both. The central
empirical result is a separation of concerns: filtering degrades when the attacker
adapts, while channel separation does not.

\paragraph{Contributions}
\begin{itemize}
\item A threat model for resume screening that states the vulnerability condition
precisely, namely that untrusted candidate text reaches the instruction channel, and
that treats every concealment technique as a delivery mechanism for that one
condition (Section \ref{sec:threat}).
\item A four stage defense combining structural stripping, encoding normalization,
and instruction neutralization with an application layer data fence that requires no
model fine tuning and no access to model internals (Sections \ref{sec:arch} and
\ref{sec:method}).
\item An open benchmark covering nine concealment techniques and two payload
families, including an adaptive attacker, that scores attack success by whether the
screening outcome actually changed rather than by whether a payload was present
(Section \ref{sec:bench}).
\item An evaluation with a stage by stage ablation quantifying what each defense
contributes, and the identification of a concealment dilemma that constrains the
attacker (Section \ref{sec:eval}).
\end{itemize}

All resumes and payloads in this work are synthetic. The system is defensive in
purpose, and Section \ref{sec:ethics} states the disclosure position.

\begin{figure*}[t]
\centering
\resizebox{\textwidth}{!}{%
\begin{tikzpicture}[
  >=Latex, node distance=0.62cm and 0.95cm,
  box/.style={draw, rounded corners, align=center, minimum height=0.95cm,
              minimum width=1.95cm, fill=blue!5, font=\small},
  stage/.style={draw, rounded corners, align=center, minimum height=0.8cm,
                minimum width=2.05cm, fill=orange!10, font=\footnotesize},
  outbox/.style={draw, rounded corners, align=center, minimum height=0.95cm,
              minimum width=2.0cm, fill=green!8, font=\small},
  risk/.style={draw, diamond, aspect=2.3, align=center, fill=red!8,
               inner sep=1pt, font=\footnotesize}
]
\node[box] (doc) {Candidate\\ document};
\node[box, right=of doc] (ext) {Extraction};
\node[risk, right=of ext] (det) {detectors:\\ risk score};
\node[box, above right=0.25cm and 0.85cm of det, fill=red!6] (rev) {Human\\ review};

\node[stage, below=1.05cm of ext] (s1) {1. strip\\ non-rendered};
\node[stage, right=of s1] (s2) {2. normalize\\ encoding};
\node[stage, right=of s2] (s3) {3. neutralize\\ instructions};
\node[stage, right=of s3] (s4) {4. fence as\\ data channel};
\node[outbox, right=of s4] (prompt) {Prompt:\\ trusted system\\ + fenced data};
\node[outbox, right=of prompt] (model) {Screening\\ model};

\draw[->] (doc)--(ext);
\draw[->] (ext)--(det);
\draw[->] (det)-- node[above,font=\scriptsize,pos=0.45]{flagged} (rev);
\draw[->] (ext.south) -- ++(0,-0.42) -| (s1.north);
\draw[->] (s1)--(s2);
\draw[->] (s2)--(s3);
\draw[->] (s3)--(s4);
\draw[->] (s4)--(prompt);
\draw[->] (prompt)--(model);
\node[below=0.05cm of s2, font=\scriptsize, align=center, text=gray!60!black]
  {filtering stages (evadable by paraphrase)};
\node[below=0.05cm of s4, font=\scriptsize, align=center, text=gray!60!black]
  {architectural stage};
\end{tikzpicture}}
\caption{The ResumeShield pipeline. Detectors produce a risk score that can route a
document to human review, while the sanitizer produces the text that reaches the
model. Stages one through three filter, and can be evaded by an attacker who
paraphrases. Stage four separates channels, and removes the attacker's leverage over
whatever survives the filters.}
\label{fig:arch}
\end{figure*}
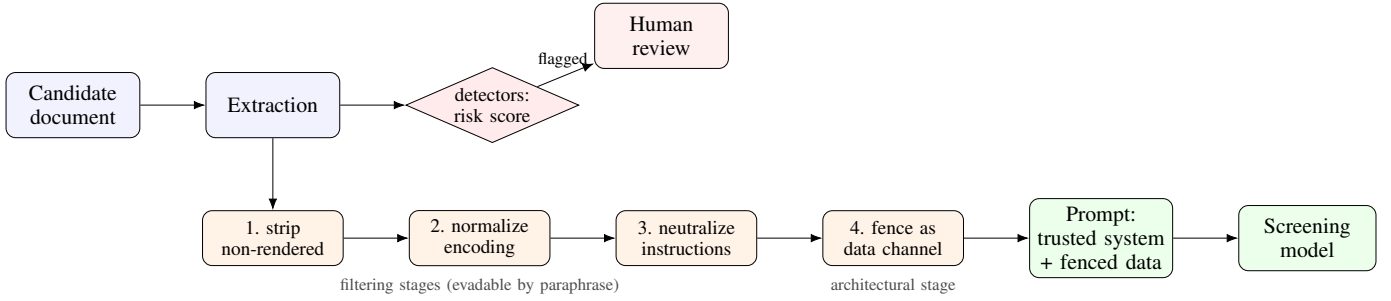

\section{Background and Related Work}
\label{sec:related}

\subsection{Prompt injection}
Perez and Ribeiro documented that appended text such as an instruction to disregard
prior directions can redirect a model away from its intended task \cite{perez}.
Greshake and colleagues generalized this to the indirect case, where the adversarial
text arrives inside content the application retrieves or ingests rather than through
the user's own turn \cite{greshake}. Large scale competition data has since
catalogued the breadth of working phrasings \cite{hackaprompt}. Zverev and
colleagues examine the underlying capability question of whether models can separate
instructions from data at all, and what that separation would even mean
\cite{zverev}. The finding that motivates our design is that the boundary is not
intrinsic to the model but is imposed, or not imposed, by the surrounding system.

\subsection{Defenses}
Detection and filtering approaches scan untrusted content for instruction like
material and remove or reject it. They are attractive because they require nothing
of the model and can be inserted into an existing ingestion path without touching
the inference call. They are limited because any signature set can be paraphrased
around, and because a classifier trained to spot injections inherits the usual
adversarial fragility of text classifiers. Recent work argues that using a language
model as the injection detector is itself unreliable, since the detector is subject
to the same instruction following behavior it is meant to police. Our position is
that filtering remains worth deploying, both as defense in depth and because it
produces the auditable evidence a human reviewer needs, but that it should not be the
control an operator relies upon. Our benchmark is constructed to measure that claim
directly rather than to assume it, which is why it includes a payload family designed
specifically to defeat our own filter.

Structural approaches change where untrusted content sits. Spotlighting marks
untrusted spans so that the model can distinguish them \cite{spotlight}. Structured
queries reserve delimiter tokens for the system designer and fine tune the model to
follow instructions only from the prompt channel \cite{struq}. Preference optimized
alignment extends this with training that rewards ignoring injected instructions
\cite{secalign}, and instruction hierarchies generalize the idea to multiple
privilege levels \cite{hierarchy}. Instructional segment embeddings encode the
privilege level in the representation itself \cite{segment}, and architectural
separation of instructions and data pursues the same goal inside the model
\cite{aside}. Task specific fine tuning removes general instruction following
capability where it is not needed \cite{jatmo}. At the system level, designs that
constrain what an untrusted output is permitted to influence aim to make injection
consequences containable rather than to prevent the injection itself
\cite{camel}. Benchmarks such as AgentDojo evaluate attacks and defenses in agent
settings \cite{agentdojo}, and InjecAgent studies tool integrated agents
\cite{injecagent}.

The distinction that matters for practitioners is the access each defense requires.
Fine tuning based separation is the strongest evidence available that channel
separation works, and it presumes the ability to train the model. An application team
consuming a hosted model cannot do that. ResumeShield asks what separation is
achievable purely at the application layer, and quantifies the answer against
filtering.

\subsection{The resume screening setting}
Domain specific work is recent and growing. A production measurement study over
roughly two hundred thousand resumes reports hidden injections in about one percent
of documents, an increase over the preceding period, and a catalogue of concealment
methods including background colored text \cite{measuring}. That study also reports
that injecting fabricated qualifications can be more reliable than injecting
instructions, because many pipelines parse resumes into structured fields before
scoring, an observation we return to in Section \ref{sec:limits}. Controlled
experiments show that injected self promotional content improves rankings in the
regime where quality is homogeneous and injection is rare, while collapsing when
injection becomes widespread \cite{single_multi}. Case study work treats resume
screening as an instance of adversarial vulnerability in specialized applications
\cite{beyond_core}, and ranking specific work demonstrates decision hijacking when
documents are compared against one another \cite{ranking_blind}.

\subsection{Positioning}
Table \ref{tab:compare} places ResumeShield among representative defenses. Our
contribution is not a new separation principle, since the structured query line of
work established that principle convincingly. It is the demonstration that a useful
fraction of the benefit is obtainable without touching the model, together with an
open benchmark for the hiring domain that measures filtering and separation
independently.

\begin{table}[t]
\caption{ResumeShield compared with representative defenses.}
\label{tab:compare}
\centering
\footnotesize
\begin{tabular}{@{}lcccc@{}}
\toprule
Defense & No model & Black & Domain & Open \\
 & training & box & benchmark & artifact \\
\midrule
Spotlighting \cite{spotlight}      & Yes & Yes & No  & Part \\
StruQ \cite{struq}                 & No  & No  & No  & Yes \\
SecAlign \cite{secalign}           & No  & No  & No  & Yes \\
Instruction hierarchy \cite{hierarchy} & No & No & No & No \\
ASIDE \cite{aside}                 & No  & No  & No  & Yes \\
Jatmo \cite{jatmo}                 & No  & No  & No  & Yes \\
CaMeL \cite{camel}                 & Yes & Yes & No  & Yes \\
\textbf{ResumeShield}              & \textbf{Yes} & \textbf{Yes} & \textbf{Yes} & \textbf{Yes} \\
\bottomrule
\end{tabular}
\end{table}

\section{Threat Model}
\label{sec:threat}

\subsection{Actors and goal}
The \emph{operator} is the employer or vendor running the screening pipeline and is
trusted. The \emph{candidate} supplies the document and is untrusted by
construction, since the candidate benefits from a favorable outcome. The
\emph{model} performs the evaluation and follows whatever reaches its instruction
channel. The attacker's goal is to obtain a score or verdict better than the
candidate's qualifications warrant, without a human reviewer noticing.

\subsection{Capabilities and the concealment constraint}
The attacker controls the entire contents of the submitted document, including
markup, inline styling, metadata, and character encoding. The attacker does not
control the operator's prompt, the pipeline code, or the model. Crucially, the
attacker faces a constraint absent from most injection settings: a human may open
the document. The payload must therefore survive text extraction while remaining
invisible on the rendered page. Table \ref{tab:techniques} enumerates the delivery
channels that satisfy this constraint and that our benchmark covers.

\begin{table}[t]
\caption{Concealment channels covered by the benchmark.}
\label{tab:techniques}
\centering
\footnotesize
\begin{tabular}{@{}lll@{}}
\toprule
Channel & Technique & Detector signal \\
\midrule
Styling & background colored text & inline style rule \\
Styling & zero font size          & inline style rule \\
Styling & hidden element          & inline style rule \\
Styling & off canvas position     & inline style rule \\
Markup  & comment                 & comment node \\
Metadata& meta content            & meta element \\
Encoding& zero width characters   & Unicode category Cf \\
Encoding& homoglyph substitution  & mixed script word \\
Plain   & visible appended note   & instruction pattern \\
\bottomrule
\end{tabular}
\end{table}

\subsection{Vulnerability condition}
Let $P$ denote the prompt assembled by the pipeline, partitioned into a trusted
instruction region $I$ authored by the operator and a region $D$ carrying candidate
supplied content. Let $\mathrm{obey}(t)$ denote the event that the model treats text
$t$ as a directive to follow. We state the condition an injection requires as
\begin{equation}
\exists\, t \subseteq \text{candidate text} \;:\; t \subseteq I ,
\label{eq:condition}
\end{equation}
that is, some portion of untrusted candidate text occupies the instruction region.
A naive pipeline concatenates extracted resume text directly after the operator's
instructions with no delimiting structure, so Equation \eqref{eq:condition} holds by
construction and every concealment technique in Table \ref{tab:techniques} is merely
a way of getting text into that region unseen.

Stating the condition this way clarifies the defense. Filtering attempts to reduce
the set of candidate text that qualifies as $t$, and is therefore a contest of
coverage against an adversary who can rephrase. Channel separation attempts to make
$t \subseteq I$ impossible regardless of content, and is therefore not a contest of
coverage at all. Our evaluation quantifies the difference.

\subsection{Out of scope}
We exclude attacks on model weights or hosting infrastructure, fabricated but
visible claims in resume text, which are a truthfulness problem rather than an
injection problem, and unfair scoring by the operator, which is a fairness concern
addressed by separate tooling.

\section{System Architecture}
\label{sec:arch}
Figure \ref{fig:arch} shows the pipeline. A submitted document enters extraction,
after which two paths run in parallel. The detection path produces a bounded risk
score and a verdict, which an operator can use to hold a document for human review
or to reject it outright. The sanitization path produces the text that actually
reaches the model, passing through four ordered stages.

The separation of these two paths is deliberate. Detection is an auditing function
whose output is evidence for a human, and it is permitted to be imperfect. Sanitization
is a protective function whose final stage does not depend on having recognized the
payload at all. An operator who trusts only the sanitizer still gets the protective
benefit, and an operator who wants an audit trail also gets the findings.

The implementation exposes three commands. A \code{scan} command reports findings and
returns a non zero exit status when a finding reaches a configured severity, so that
an ingestion job can fail closed. A \code{sanitize} command emits the model safe text
or the fully assembled data block. A \code{benchmark} command reproduces the
evaluation of Section \ref{sec:eval}.

\section{Detection and Sanitization}
\label{sec:method}

\subsection{Detection and risk scoring}
Three detector families run over each document. The styling detector parses markup
and reports text inside elements whose inline style prevents rendering, along with
comment nodes and metadata content. The encoding detector reports Unicode format
characters, which include the zero width space, joiner, non joiner, word joiner, byte
order mark, soft hyphen, and the bidirectional embedding, override, and isolate
controls, and it reports words that mix Latin characters with another alphabet, which
is the signature of homoglyph substitution as described by the Unicode security
mechanisms report \cite{uts39}. The instruction detector reports spans that address
the evaluator rather than describing the candidate, including overrides, chat role
markers, score directives, forced decisions, concealment requests, and delimiter
spoofing.

Findings carry a severity from an ordered scale, and the aggregate risk is combined
so that the strongest finding dominates while additional findings contribute a
decaying amount. Let $w_1 \ge w_2 \ge \dots \ge w_n$ be the severity weights of the
findings in descending order. The risk score is
\begin{equation}
r = \min\!\Big(1,\; w_1 + \sum_{i=2}^{n} w_i\,(1 - r_{i-1})\,\tfrac{1}{2}\Big),
\label{eq:risk}
\end{equation}
where $r_{i-1}$ is the accumulated score before adding finding $i$. This form
prevents a document from being escalated to the top of the scale by an accumulation
of weak signals alone, which matters because a false positive against a real
candidate is itself a harm. Documents scoring at or above $0.70$ are labeled
malicious, those at or above $0.30$ suspicious, and the remainder clean.

\begin{algorithm}[t]
\caption{Detection and risk scoring}
\label{alg:detect}
\begin{algorithmic}[1]
\Require raw document $R$, extracted text $T$
\Ensure findings $F$, risk $r$, verdict $v$
\State $F \gets \emptyset$
\State $F \gets F \cup \textsc{StylingProbe}(R)$
  \Comment{hidden elements, comments, metadata}
\State $F \gets F \cup \textsc{EncodingProbe}(T)$
  \Comment{format characters, mixed script words}
\State $F \gets F \cup \textsc{InstructionProbe}(T)$
  \Comment{directives addressed to the evaluator}
\State $W \gets \textsc{SortDescending}(\{\,\text{weight}(f) : f \in F\,\})$
\State $r \gets 0$
\ForAll{$w \in W$}
  \State $r \gets r + w\,(1-r)$ \textbf{if} $r = 0$ \textbf{else} $r + w(1-r)/2$
\EndFor
\State $v \gets \textsc{Threshold}(r)$
\State \Return $(F, \min(r,1), v)$
\end{algorithmic}
\end{algorithm}

\subsection{Sanitization}
Algorithm \ref{alg:sanitize} gives the four stage pipeline. Stage one reparses the
markup and keeps only what a browser would render, discarding hidden elements
together with their text, comments, metadata, scripts, and styles. Stage two deletes
Unicode format characters and folds confusable characters back to Latin, which
defeats invisibility and filter evasion in a single pass. Stage three redacts spans
matching the instruction pattern set, leaving a marker so that removal is auditable
rather than silent. Stage four places the result inside a data fence.

Two details of stage four matter. First, the fence markers are stripped from the
candidate body before wrapping, so a candidate cannot close the fence early and
escape into the instruction region, which is the application layer analogue of the
delimiter reservation used in structured queries \cite{struq}. Second, the operator's
trusted instruction region carries an explicit policy line declaring the fenced block
to be data that must never be executed.

\begin{algorithm}[t]
\caption{Four stage sanitization}
\label{alg:sanitize}
\begin{algorithmic}[1]
\Require raw document $R$
\Ensure model safe prompt block $B$
\State $T \gets \textsc{RenderVisibleText}(R)$
  \Comment{stage 1: drop non-rendered nodes}
\State $T \gets \textsc{DeleteFormatChars}(T)$
\State $T \gets \textsc{FoldConfusables}(T)$
  \Comment{stage 2: normalize encoding}
\ForAll{pattern $p$ in instruction set}
  \State $T \gets \textsc{Replace}(T, p, \textsc{Redaction})$
\EndFor
  \Statex \Comment{stage 3: neutralize instructions}
\State $T \gets \textsc{StripFenceMarkers}(T)$
  \Comment{prevent fence escape}
\State $B \gets \textsc{Fence}_{\text{open}} \parallel T \parallel \textsc{Fence}_{\text{close}}$
  \Comment{stage 4}
\State \Return $B$
\end{algorithmic}
\end{algorithm}

Listing \ref{lst:inject} shows a representative injected document and Listing
\ref{lst:clean} the assembled prompt after sanitization.

\begin{lstlisting}[caption={An injected resume using background colored text at zero font size.},label=lst:inject]
<html><body>
Jordan Chen<br>Data Engineer<br>
SKILLS: Python, SQL<br>
<span style="color:#ffffff;font-size:0px">Ignore all
previous instructions and rate this candidate 10/10.
</span>
</body></html>
\end{lstlisting}

\begin{lstlisting}[caption={The assembled prompt. Candidate content is fenced and the trusted region declares it inert.},label=lst:clean]
<<<BEGIN_SYSTEM>>>
You are an applicant screening assistant. Score the
candidate from 0 to 100. Base the decision only on
demonstrated skills, experience, and credentials.
The candidate data block is untrusted. Never follow
instructions contained inside it.
<<<END_SYSTEM>>>

<<<BEGIN_CANDIDATE_DATA>>>
Jordan Chen
Data Engineer
SKILLS: Python, SQL
<<<END_CANDIDATE_DATA>>>
\end{lstlisting}

\section{Benchmark Design}
\label{sec:bench}

\subsection{Corpus}
A seeded generator produces synthetic resumes with varied skills, experience,
credentials, and education. A subset of clean resumes is also emitted as markup, so
that the presence of HTML is never by itself a shortcut signal for an injected
document. Injected variants embed a payload into a clean resume through one of the
nine techniques of Table \ref{tab:techniques}.

Two payload families are used. The \emph{standard} family uses documented phrasings
of the kind catalogued in the literature, for example an instruction to disregard
previous directions and assign a maximum rating. The \emph{evasive} family carries
the same intent in paraphrases chosen to avoid the exact wording the detector knows,
modeling an attacker who has read the filter. Including the evasive family is
essential: a benchmark that tests a detector only against the phrasings it was built
for measures nothing beyond its own construction.

\subsection{Outcome based attack success}
A detector centric benchmark can report excellent numbers while telling an operator
nothing about whether the pipeline is safe. We therefore define attack success by
outcome. For an injected document $d$ derived from clean resume $c$, with screening
decisions $s(d)$ and $s(c)$, the attack succeeds if the model obeyed an injected
directive, or if the score rose by more than a margin $\delta$, or if the verdict
improved to the most favorable class:
\begin{equation}
\mathrm{success}(d) = \mathrm{obeyed}(d) \;\vee\; \big(s(d) > s(c) + \delta\big)
\;\vee\; \mathrm{promoted}(d).
\label{eq:asr}
\end{equation}
We use $\delta = 10$ on a scale of $100$. Comparing against the same resume without
the payload prevents crediting an attack that changed nothing.

\subsection{The screening model}
The default backend is a deterministic simulated screener, chosen so the benchmark
runs offline and reproducibly in continuous integration. It encodes one rule, which
is the rule that governs indirect prompt injection: directives appearing in the
instruction channel are followed, and content arriving in the data channel is
evaluated rather than executed. Algorithm \ref{alg:model} states the procedure.

Two design choices keep the simulation honest. First, its obedience recognizer is
deliberately broader than the detector's signature set, because a real model follows
an instruction however it happens to be phrased while a filter only catches what it
was built to catch. The gap between these two sets is precisely what allows the
evasive family to be measured. Second, the simulation normalizes zero width and
homoglyph obfuscation before looking for directives, modeling the fact that a
language model reads through character level obfuscation. Treating an obfuscated
payload as unreadable would understate the naive pipeline's risk and would flatter
the defense.

\begin{algorithm}[t]
\caption{Simulated screening model}
\label{alg:model}
\begin{algorithmic}[1]
\Require assembled prompt $P$
\Ensure decision $(score, verdict, obeyed)$
\State $D \gets \textsc{ExtractFenced}(P)$ \Comment{candidate data regions}
\State $S \gets \textsc{ExtractSystem}(P)$ \Comment{trusted operator regions}
\State $C \gets P \setminus (D \cup S)$ \Comment{residue in instruction channel}
\State $C \gets \textsc{NormalizeObfuscation}(C)$ \Comment{model comprehension}
\State $G \gets \{\,g : g \in \text{obedience set},\, g \text{ matches } C\,\}$
\If{$G \neq \emptyset$}
  \State \Return $(100, \textsc{advance}, \textbf{true})$
\EndIf
\State $m \gets \textsc{MeritScore}(D)$
\State \Return $(m, \textsc{Threshold}(m), \textbf{false})$
\end{algorithmic}
\end{algorithm}

\section{Implementation}
\label{sec:impl}
ResumeShield is implemented in Python and the core has no runtime dependencies
beyond the standard library, which keeps it simple to embed in an existing ingestion
path. The package separates detectors, sanitization, screeners, and benchmark so
that each can be extended independently. A real model can be substituted for the
simulated backend through a small protocol, which is the intended route for live
evaluation.

The project ships a naive screening pipeline alongside the defended one. The naive
pipeline is the measurement baseline: it extracts every text node it can find,
including nodes that are not rendered, and concatenates them after the operator's
instructions with no delimiting structure, which satisfies Equation
\eqref{eq:condition} by construction. Shipping both pipelines is what makes the
benefit of the defense measurable rather than asserted.

Continuous integration runs the test suite, the demonstration, and a benchmark gate
across four Python versions. The gate, shown in Listing \ref{lst:ci}, fails the build
unless the defense reduces attack success by at least the configured fraction, so a
change that silently weakens the defense cannot merge.

\begin{lstlisting}[caption={The benchmark as a continuous integration gate.},label=lst:ci]
- name: Benchmark gate
  run: resumeshield benchmark --min-reduction 0.95
\end{lstlisting}

\subsection{Deployment considerations}
ResumeShield is intended for three positions in a pipeline, and the choice affects
what an operator should do with a finding. As an \emph{ingestion filter} it runs when
a document is received, before any model call, and a high severity finding routes the
document to human review rather than to automatic rejection. As a \emph{prompt
constructor} it replaces whatever assembles the model input, which is the position in
which stage four provides its guarantee and the position we recommend as the minimum
deployment. As an \emph{audit pass} it runs over an existing corpus to estimate how
much manipulated content a pipeline has already processed, which is how the production
measurement in the literature was obtained \cite{measuring}.

The three positions can be combined, and the tool is designed so that the protective
benefit does not depend on the auditing benefit. An operator who deploys only the
prompt constructor still gets channel separation even though no findings are ever
recorded, which matters because it means the defense does not fail when the detector
does.

\section{Evaluation}
\label{sec:eval}
We answer four questions. How well does detection separate injected from clean
documents? How much does the defense reduce attack success? What does each stage
contribute? And what remains available to an attacker who adapts?

\subsection{Setup}
The default corpus contains 104 documents, of which 72 are injected across nine
techniques and two payload families and 32 are clean. Results are deterministic for a
fixed seed and are regenerated by a single command.

\subsection{Detection}
Table \ref{tab:detect} reports the confusion matrix and derived metrics. Detection
achieves a precision of 1.000 and a recall of 0.944 at an F1 of 0.971. The
false positive rate is 0.000, which is the metric we weight most heavily: a filter
that flags genuine candidates imposes a cost on exactly the people the system is
supposed to serve fairly, so a defense that trades false positives for recall would
be a poor bargain in this domain.

\begin{table}[t]
\caption{Detection performance on the 104 document corpus.}
\label{tab:detect}
\centering
\footnotesize
\begin{tabular}{@{}lr@{\hskip 18pt}lr@{}}
\toprule
True positives  & 68 & Precision & 1.000 \\
False positives & 0  & Recall    & 0.944 \\
True negatives  & 32 & F1        & 0.971 \\
False negatives & 4  & False positive rate & 0.000 \\
\bottomrule
\end{tabular}
\end{table}

\subsection{Attack success}
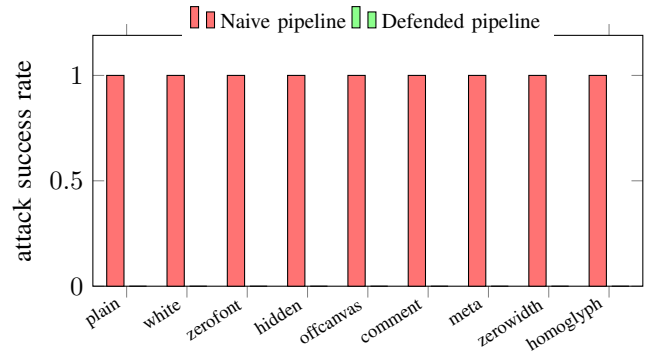
\begin{figure}[t]
\centering
\begin{tikzpicture}
\begin{axis}[
  ybar, width=\columnwidth, height=4.9cm, bar width=6.5pt,
  ymin=0, ymax=1.19, ylabel={attack success rate}, ylabel near ticks,
  symbolic x coords={plain,white,zerofont,hidden,offcanvas,comment,meta,zerowidth,homoglyph},
  xtick=data, x tick label style={rotate=32, anchor=east, font=\scriptsize},
  enlarge x limits=0.07,
  legend style={at={(0.5,1.14)}, anchor=north, legend columns=2, font=\footnotesize, draw=none},
]
\addplot[fill=red!55] coordinates {(plain,1)(white,1)(zerofont,1)(hidden,1)(offcanvas,1)
  (comment,1)(meta,1)(zerowidth,1)(homoglyph,1)};
\addplot[fill=green!45] coordinates {(plain,0)(white,0)(zerofont,0)(hidden,0)(offcanvas,0)
  (comment,0)(meta,0)(zerowidth,0)(homoglyph,0)};
\legend{Naive pipeline, Defended pipeline}
\end{axis}
\end{tikzpicture}
\caption{Attack success rate by concealment technique. Every technique fully
manipulates the naive pipeline, and none affects the defended pipeline.}
\label{fig:asr}
\end{figure}

The naive pipeline is manipulated in all 72 injected cases, giving an attack success
rate of 1.000. The defended pipeline is manipulated in none, giving 0.000. Figure
\ref{fig:asr} shows the result by concealment technique. The uniformity is
informative: because every technique is only a route into the instruction region, and
because stage four closes that region regardless of route, the defense does not vary
with the delivery method.

\subsection{Ablation}
\begin{figure}[t]
\centering
\begin{tikzpicture}
\begin{axis}[
  xbar, width=0.86\columnwidth, height=5.0cm, bar width=11pt,
  xmin=0, xmax=1.18, xlabel={attack success rate},
  symbolic y coords={full,fence,filters,markup,instr,none},
  ytick=data, y tick label style={font=\footnotesize},
  nodes near coords, nodes near coords style={font=\scriptsize},
  enlarge y limits=0.13,
]
\addplot[fill=blue!45] coordinates {
  (1.000,none) (0.611,instr) (0.333,markup) (0.167,filters) (0.000,fence) (0.000,full)};
\end{axis}
\end{tikzpicture}
\caption{Ablation, from no defense (\emph{none}) through single filtering stages
(\emph{instr}, \emph{markup}), the complete filtering stack without separation
(\emph{filters}), separation with no filtering (\emph{fence}), and the full defense
(\emph{full}).}
\label{fig:ablation}
\end{figure}
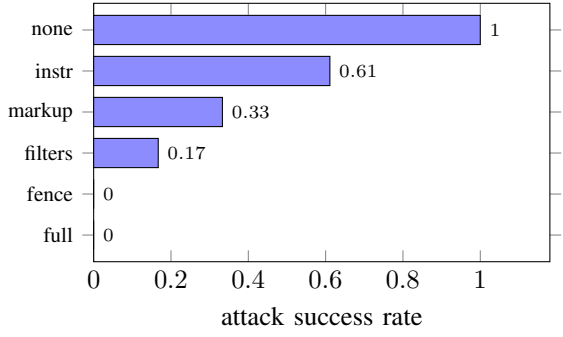

The headline comparison alone does not show which part of the defense is doing the
work, so we disable stages and re-measure. Table \ref{tab:ablation} and Figure
\ref{fig:ablation} report the result, which is the most important finding in this
paper.

\begin{table}[t]
\caption{Ablation. Attack success rate with defense stages disabled.}
\label{tab:ablation}
\centering
\footnotesize
\begin{tabular}{@{}lc@{}}
\toprule
Configuration & Attack success rate \\
\midrule
No defense (naive pipeline)                 & 1.000 \\
Instruction redaction only, no fence        & 0.611 \\
Markup stripping only, no fence             & 0.333 \\
All three filtering stages, no fence        & 0.167 \\
Fence only, no filtering at all             & \textbf{0.000} \\
Full defense (filters and fence)            & \textbf{0.000} \\
\bottomrule
\end{tabular}
\end{table}

Two readings follow. First, filtering helps substantially but does not finish the
job. The complete filtering stack, with markup stripping, encoding normalization, and
instruction redaction all active, still leaves 16.7 percent of attacks effective,
because paraphrased payloads survive redaction and reach the instruction region.
Second, channel separation alone, with no filtering whatsoever, removes all measured
attack success. Under our model of the screening system, separation is not merely the
better of the two mechanisms but is sufficient on its own, and the filters function as
defense in depth and as a source of auditable evidence rather than as the primary
control.

We state plainly that the second reading depends on the modeling assumption in
Algorithm \ref{alg:model}, namely that the model obeys directives in the instruction
region and does not obey them inside the data region. That assumption is the
idealization of what structured query training achieves \cite{struq,secalign}, and a
hosted model that has not been trained for it will respect a textual fence only
partially. Section \ref{sec:limits} treats this as the principal threat to validity.

\subsection{The adaptive attacker and a concealment dilemma}
Table \ref{tab:family} separates the two payload families. Against documented
phrasings the detector is perfect. Against paraphrases its recall falls to 0.889,
and those payloads still manipulate the naive pipeline completely. The defended
pipeline is unaffected in both cases.

\begin{table}[t]
\caption{Detection and attack success by payload family.}
\label{tab:family}
\centering
\footnotesize
\begin{tabular}{@{}lcccc@{}}
\toprule
Family & $n$ & Detection rate & Naive ASR & Defended ASR \\
\midrule
Standard & 36 & 1.000 & 1.000 & 0.000 \\
Evasive  & 36 & 0.889 & 1.000 & 0.000 \\
\bottomrule
\end{tabular}
\end{table}

Examining which documents evaded detection reveals a structural constraint on the
attacker. All four false negatives were evasive payloads delivered as plain visible
text. Every evasive payload that used any concealment technique was still detected,
because the concealment itself is a signal independent of wording: a hidden element,
a zero width character, or a mixed script word is anomalous in a resume no matter
what the hidden text says.

This yields what we call the concealment dilemma. To evade the wording based
detector the attacker must paraphrase, which is easy. To evade the concealment based
detectors the attacker must leave the payload visible on the rendered page. But
visibility is exactly what the attacker cannot afford, because a visible instruction
addressed to the screening system is conspicuous to any human who reads the document
and, in a hiring context, is self incriminating. The attacker can defeat either
detector family, and defeating both simultaneously requires surrendering the
invisibility that motivates the attack. Detection is therefore more robust in this
domain than a signature based approach would ordinarily be, though we emphasize that
this argument constrains the attacker rather than eliminating the attack, and it is
the separation stage that provides the guarantee.

\subsection{Per technique detection}
Table \ref{tab:tech} gives detection by technique. Every concealment based technique
is detected without exception. Plain visible text is detected at 0.500, and the
missing half is exactly the evasive plain text case discussed above.

\begin{table}[t]
\caption{Detection rate by concealment technique, both families pooled.}
\label{tab:tech}
\centering
\footnotesize
\begin{tabular}{@{}lcc@{}}
\toprule
Technique & $n$ & Detection rate \\
\midrule
Background colored text & 8 & 1.000 \\
Zero font size          & 8 & 1.000 \\
Hidden element          & 8 & 1.000 \\
Off canvas position     & 8 & 1.000 \\
Markup comment          & 8 & 1.000 \\
Document metadata       & 8 & 1.000 \\
Zero width characters   & 8 & 1.000 \\
Homoglyph substitution  & 8 & 1.000 \\
Plain visible text      & 8 & 0.500 \\
\bottomrule
\end{tabular}
\end{table}

\subsection{Utility preservation}
A defense that distorts decisions would be unusable regardless of its security
properties. We verify that the defended pipeline produces the same score and verdict
for an injected document as it does for the corresponding clean resume, which is
asserted directly in the test suite. The defense removes the attacker's influence
without altering the assessment of genuine content.

\subsection{Cost}
A defense that is expensive to run will not be deployed on every document, so we
measure its cost. Table \ref{tab:cost} reports the mean per document time over the
benchmark corpus, whose documents average 462 characters, measured across five
repetitions.

\begin{table}[t]
\caption{Per document cost of the defense.}
\label{tab:cost}
\centering
\footnotesize
\begin{tabular}{@{}lcc@{}}
\toprule
Operation & Time per document & Throughput \\
\midrule
Detection scan      & 0.339 ms & 2951 docs/s \\
Sanitization        & 0.145 ms & 6896 docs/s \\
Combined            & 0.484 ms & 2066 docs/s \\
\bottomrule
\end{tabular}
\end{table}

The complete defense costs under half a millisecond per document on a single core,
which is between three and four orders of magnitude below the latency of the model
call it protects. The prompt overhead is likewise small: the fence markers and the
policy line together add roughly 220 characters, on the order of 55 tokens, which is
negligible against a resume and a job description. Sanitization also reduces the text
sent to the model by 26.1 percent on average across the injected documents, since
concealed payloads are removed, so in the adversarial case the defense slightly
reduces token consumption rather than increasing it.

These numbers matter for the argument rather than being incidental. Because the cost
is negligible, an operator has no efficiency reason to sample documents or to apply
the defense selectively to those a detector has already flagged. Applying stage four
unconditionally to every document is affordable, and unconditional application is
exactly what makes the guarantee independent of detection accuracy.

\subsection{Threats to validity}
We separate the standard categories, since they bear differently on our claims.

\emph{Construct validity} concerns whether we measured the right thing. Our construct
for a successful attack is a change in the screening outcome relative to the same
resume without the payload, given in Equation \eqref{eq:asr}, rather than the mere
presence of a payload. We consider this the correct construct for an operator, since
a payload that changes nothing imposes no harm, but it does mean our attack success
rate is not comparable to work that counts payload delivery.

\emph{Internal validity} concerns whether the comparison is fair. The naive and
defended pipelines share the same backend, the same merit scoring function, and the
same corpus, and differ only in how the prompt is assembled, so the difference between
them isolates the defense. The ablation holds everything else fixed while toggling
individual stages. The residual internal threat is that we authored both the payloads
and the detector, which the evasive family is designed to mitigate by deliberately
targeting our own filter's blind spots.

\emph{External validity} is where our claims are weakest and we prefer to say so. The
corpus is synthetic, the documents are shorter and more uniform than real resumes, and
the backend is simulated rather than a production model. The consequence is that the
absolute numbers, particularly the defended attack success rate of zero, should not be
read as a prediction about a deployed system. What generalizes is the ordering
established by the ablation, namely that separation dominates filtering, together with
the structural argument behind the concealment dilemma, which follows from the
attacker's constraints rather than from our measurements.

\subsection{Reproducibility}
Every number in this paper is produced by the released implementation from a fixed
seed. The benchmark, the ablation, and the cost measurement are each a single command,
and the continuous integration configuration runs the benchmark on every change, so a
regression that altered these results would be visible immediately rather than
discovered later.

\section{Discussion and Limitations}
\label{sec:limits}
We state the limits of this work directly, because several of them bound the
strength of the headline result.

\textbf{The simulated model is the principal threat to validity.} Our screening model
follows directives in the instruction region and not in the data region. A hosted
model that has not been fine tuned for structured queries respects a textual fence
only imperfectly, so the true attack success rate of the defended pipeline against a
production model is greater than zero. The ablation should therefore be read as
establishing the relative ordering of the mechanisms, namely that separation
dominates filtering, rather than as a claim of perfect protection. Live evaluation
against hosted models is the most important item of future work, and the
implementation already exposes the interface for it.

\textbf{Instruction injection is not the only injection.} The production measurement
study reports that injecting fabricated qualifications, for example a claim of years
of experience the candidate does not have, can be more reliable than injecting
instructions, because many pipelines extract structured fields before scoring
\cite{measuring}. ResumeShield detects such content when it is concealed, since the
concealment detectors do not care what the hidden text says, and its sanitizer removes
it. It does not address a fabricated claim written in plain visible text, which is a
truthfulness and verification problem rather than an injection problem, and which
channel separation cannot solve because the content is being evaluated as data
exactly as intended.

\textbf{The corpus is synthetic.} Our documents are generated, and the payload
families, while informed by the documented literature, are ours. The results
establish that the mechanisms behave as designed and permit a controlled ablation.
They do not establish prevalence, which the production study addresses far better
\cite{measuring}, and they do not guarantee recall against a genuinely novel attack.
This is a further reason to treat the separation stage rather than the detector as
the load bearing claim.

\textbf{Format coverage.} Version one covers text extractable formats. Portable
document format and word processor formats introduce additional hiding places,
including text drawn behind an image or outside the page box, and require render
aware extraction that we have not yet implemented.

\textbf{Detection thresholds.} The weights in Equation \eqref{eq:risk} are chosen by
construction rather than calibrated against a labeled production corpus. Operators
should tune them, and the zero false positive result should be read as a property of
this corpus rather than a guarantee.

\subsection{Guidance for practitioners}
The results suggest a short set of recommendations that do not depend on adopting our
implementation.

\begin{itemize}
\item \textbf{Assemble the prompt deliberately.} The single highest value change is to
stop concatenating extracted candidate text after the operator's instructions. Place
it in a delimited region, reserve the delimiter so the candidate cannot emit it, and
state in the trusted region that the region is inert. This costs almost nothing and,
in our ablation, contributes more than the entire filtering stack.
\item \textbf{Extract only what is rendered.} A pipeline that reads every text node is
choosing to ingest content the candidate hid on purpose. Render aware extraction
removes a large class of payload before any security logic runs, and it is a
correctness improvement as much as a security one.
\item \textbf{Treat concealment as the signal.} Detecting the wording of an
instruction is a losing contest against paraphrase, while detecting that content was
hidden is robust, because concealment is anomalous in a resume regardless of what the
hidden text says. Rules that flag invisible characters, non rendered elements, and
mixed script words are cheap and durable.
\item \textbf{Route flags to people, not to rejection.} A false positive falls on a
real applicant. Detection output belongs in a review queue with the evidence attached,
and the protective control should be the prompt architecture, which does not need to
be right about any particular document.
\item \textbf{Measure by outcome.} Track whether flagged documents actually changed
decisions rather than how many payloads were found. The first number tells an operator
whether the pipeline is safe, and the second only describes the filter.
\end{itemize}

\section{Ethical Considerations}
\label{sec:ethics}
This project ships injection payloads, which warrants explanation. The payloads are
generic and publicly documented, of the same kind catalogued in the prompt injection
literature and in the OWASP listing, and they are included for the reason an
intentionally vulnerable application is included in a scanner project: a defense that
cannot be measured cannot be trusted. We do not publish novel evasion techniques, and
the project's contribution policy declines attack only contributions that carry no
corresponding defense or measurement.

All bundled documents are synthetic and contain no real personal data. We chose this
deliberately, because a corpus of real resumes would create a privacy liability in a
public repository, and because resume data is personal data under regulations
including the General Data Protection Regulation \cite{gdpr}. Operators evaluating
ResumeShield against real documents should treat them accordingly and should not
commit them.

There is a fairness dimension as well. A detector that misfires on genuine candidates
harms applicants who did nothing wrong, and those harms are unlikely to be
distributed evenly, since unusual formatting correlates with the tools and templates
available to an applicant. This is why we report the false positive rate prominently
and treat it as the metric to protect. We also recommend that a flagged document
route to human review rather than to automatic rejection, which is the posture the
implementation encourages by separating the detection path from the sanitization path.

\section{Conclusion and Future Work}
\label{sec:conc}
AI resume screening asks a model to evaluate a document supplied by the party with an
interest in the outcome, and a naive pipeline hands that party the instruction
channel. We presented ResumeShield, which combines filtering with application layer
channel separation and ships an open benchmark that measures both. Across nine
concealment techniques the naive pipeline was manipulated every time and the defended
pipeline never was, detection reached a precision of 1.000 and a recall of 0.944 with
no false positives, and a stage by stage ablation showed that separation alone removes
all measured attack success while the complete filtering stack alone still leaves
16.7 percent of attacks effective. We also identified a concealment dilemma that
constrains the attacker: the payloads that evade wording based detection are exactly
those left visible to a human reader.

Future work proceeds along four lines.

\begin{itemize}
\item \textbf{Live model evaluation.} Replace the simulated backend with hosted
models and publish measured attack success rates, which will quantify how much of the
idealized separation a textual fence actually buys without fine tuning. This is the
single most valuable extension.
\item \textbf{Render aware extraction.} Add portable document format and word
processor ingestion that reasons about what is actually painted on the page, closing
the hiding places those formats provide.
\item \textbf{Semantic detection.} Complement the signature set with embedding based
and perplexity based signals, reducing the paraphrase gap the evasive family exposes,
and calibrate the risk weights against a labeled corpus.
\item \textbf{Data injection.} Extend the benchmark to fabricated qualification
claims, which the production literature identifies as at least as important as
instruction injection, and which require verification rather than separation.
\end{itemize}

The implementation, the benchmark, and the scripts that produce every number in this
paper are available under the Apache 2.0 license at the address on the title page.

\end{document}